\documentclass[twoside]{ilcws08}
\usepackage[latin1]{inputenc}
\usepackage[dvips]{graphicx,epsfig,color}
\usepackage{wrapfig,rotating}
\usepackage{amssymb,amsmath,array}

\begin{document}
\title{
Status of the CALICE DAQ system} 
\author{Valeria Bartsch$^1$ (for the CALICE-UK collaboration)
\vspace{.3cm}\\
1- University College London, Department of Physics and Astronomy \\
Gower Street, London, WC1E 6BT, United Kingdom
}

\maketitle

\begin{abstract}
A data acquisition (DAQ) system is described which will be used for the next generation 
of prototype calorimeters using particle flow algorithms for the International Linear 
Collider (ILC). The design is 
sufficiently generic and scalable such that it should have numerous applications either 
for ILC detectors or elsewhere within high energy physics in general. The DAQ system will 
be implemented using FPGAs and built using off-the-shelf components and networking hardware 
with programmable FPGAs. The software for the DAQ system is based on an existing framework, 
DOOCS, which is a server/client object-oriented system. The design philosophy, current 
status of the project and its aims are presented in this report. 
\end{abstract}

\section{Motivation and overview of the data acquisition System}

The concept of particle flow algorithms (PFA) and the beam structure of the planned 
International Linear Collider (ILC) accelerator put constraints on the data acquisition 
(DAQ) system which are described in the following. PFA is a widely accepted approach to 
improve the energy resolution at the ILC. Particle flow 
uses the high segmentation of about 1\,cm$\times$1\,cm (or smaller) in the detection 
layer of a sampling calorimeter to track particles through the calorimeter. The 
segmentation results in about 24 million readout pads in the electromagnetic calorimeter 
(ECAL) of a planned ILC detector\,\cite{TESLA TDR}. Because of 
the large number of readout channels it is necessary to minimise the cost of the 
electronics. This is achieved by using standard networking chipsets and protocols. 
Already at an early stage of the calorimeter prototypes the DAQ system has been 
designed for scalability. In the current detector plans the space for the electronics 
and cooling inside the calorimeters is very restricted. The 
DAQ system under development is aiming at a generic DAQ system, 
which will be applicable for the ECAL, analogue hadronic 
calorimeter (AHCAL) and digital hadron calorimeter (DHCAL) prototypes. At the ILC it is assumed 
that all data will be collected with no triggers 
implemented. During a bunch train, data will be stored on the detector and then read 
out during the inter-train gap of about 200\,ms. 

\begin{figure}[htb]
\begin{center}
\includegraphics[width=0.8\textwidth]{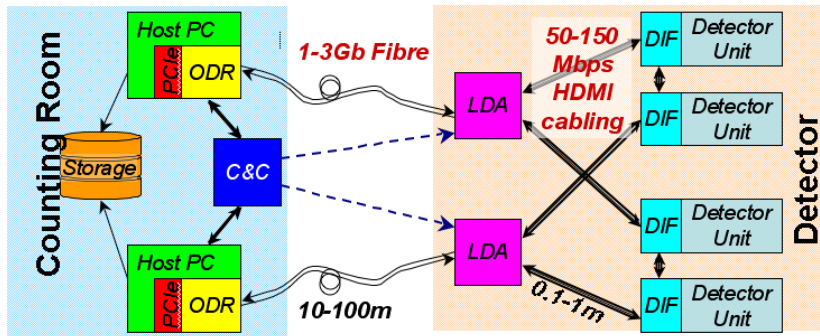}
\end{center}
\caption{An overview over the DAQ architecture}
\label{fig_daqoverview}
\end{figure}

The data are taken by the detectors on the detection slabs and digitised by the very 
front-end on-detector electronics or slabs. Then all output data are delivered to 
electronics boards at the end of the detector. Data is then sent to a concentrator 
card which collects data from many detector units before being sent off the detector on 
Gigabit optical fibres to a detector receiver in the counting room. Figure\,\ref{fig_daqoverview} gives an overview of the generic DAQ scheme.

The calorimeter  is designed such that VFE chips are 
embedded in the detector slabs. The slabs are served from the ends with a Detector 
Interface (DIF). The DIFs transmit the data to the Link-Data Aggregator (LDA) which 
is fitted on-detector and serves many DIFs via standard HDMI links. The data from the 
LDAs are fed to the Off-Detector Receiver (ODR) via the Gigabit optical cables. The 
ODR is a dedicated PCI-Express card which is hosted in a DAQ PC, being capable of saving 
the data to central storage. 

\subsection{Detector Interface Board}
The detector signals are buffered and digitised by the VFE ASICs 
situated on the detector slabs. The VFE ASICs depend on the detector used in the 
prototypes accommodating the different features of each detectors. The interface 
between the VFE and the DAQ 
system is provided by the DIF board which is located at the end of the calorimeter 
slabs. The DIF board consists of a customised part which translates the signals of the 
ASICs into more generic signals which are independent of the detector used for the 
calorimeter. The interface has been defined together with three detector groups working
on prototypes for the ECAL, AHCAL and DHCAL within the EUDET\,\cite{EUDET} project
and is common to the three different detector types. A generic part of the DIF passes 
the data on and sends configuration on to the VFE electronics. HDMI\,\cite{HDMI} is a home 
entertainment system standard with small cabling and connectors which are commercially 
available at low cost. They are rated at more than 300\,Mb/sec for data. To provide 
redundancy against link failures, a provision is made on the DIF to connect to its 
neighbour. By connecting even and odd DIFs to different LDAs, the system is protected 
for the loss of a DIF-LDA link, or even the failure of an entire LDA.

\subsection{Link Data Aggregator}
In order to accumulate the data sent from the DIF, an LDA collects the data from several 
DIFs and sends it further to the ODR.  The number of DIFs aggregated in the LDA depends 
on the detector type chosen in the detection layer. The optimal number of DIFs/LDA should take 
into account the number of available pins on the FPGA, the bandwidth per DIF and the
effective performance-price ratio to maximise bandwidth of the optical link. The current LDA 
prototype can host up to eight links to the DIFs with a small upgrade up to ten DIF connections. 
The LDA has banks of HDMI connectors for connections to the DIFs and 
a small form-factor pluggable (SFP) connector for the optical link off-detector. The 
prototype version of the LDA is built on a commercial development board with a Xilinx 
Spartan3-200 FPGA. The two sets of connectors are physically add-on boards: one providing 
the SFP and serialiser chipset for the optical link, the other hosting 
eight working HDMI connectors with clock fan-out hardware. A first 
prototype of the LDA has been produced by the company Enterpoint \,\cite{Enterpoint} and has 
been thoroughly tested.

\subsection{Off Detector Receiver}
The LDA is situated at the edge of the detector, the ODR however is located in the counting 
room connected to the on-detector DAQ system by an optical fibre. As well as receiving data 
from the detector, it sends control and configuration data to the LDA for distribution to 
the DIFs and finally sends the data off the detector to the event building. It is realised as a 
PCI-Express card and can serve up to four LDAs per 
card, with one PC hosting up to two ODRs. For the proposed DAQ system currently a 
Xilinx\,Virtex\,4 FPGA is used, thereby using a commercial FPGA board  designed by 
PLDA\,\cite{PLDA}.  At present, the data stream which will come from the detector in the 
EUDET test beam is simulated by an internal data generator in the firmware or via an on-board 
ethernet interface; this allows for a thorough debugging and optimisation of the system 
before integration with the detectors.

The user interface to the ODR card contains two parts: a customised driver and a client 
program. The former is mainly tasked with mapping card memory to the user space and providing 
direct memory access (DMA) support. The latter client retrieves data from the ODR card memory 
and stores it on the local disk. The performance of the 
prototype ODR has been investigated using an ethernet interface to provide an input data stream 
which is copied to host memory and is currently higher than 150 MB/s  writing to an array of 
disks. A higher data rate of up to 700\,MB/s can be achieved 
when writing into the host memory. The number of DMAs has been optimised for the performance. The 
ODR prototype is ready and can be optimised further in the next months. 

\section{Clock and Control Handling}
For the event building a good clock is essential because the events will be built using time 
stamps. It is understood that a machine clock will be fed into the ODRs and fanned out to the 
LDAs and DIFs. The requirements on the clock are a low jitter and a fixed latency between the 
machine clock and the clock in the DIFs. Commercial networking hardware  is not suited for 
this task as it is most efficient when it can buffer data and provides no guarantees on 
delivery times. Similarly networking hardware built into modern FPGAs suffers from varying 
latency. Therefore a clock and control (C\&C) board has been designed with links to eight ODRs. 
The C\&C logic is implemented on a board with 6U format, translating 
into a board size of 100mm\,$\times$\,160\,mm.

The clock and control module must interface with the machine and provide stand-alone signal 
and clock generation. The machine clock is expected to run between 50 and 100\,MHz. The bunch 
clock will be derived as a multiple of the machine clock. It is able to deliver fast 
asynchronous triggers. It will also receive a busy signal from the VFE. The HDMI cables used 
for the DIF to LDA link will be reused here for sending control data. The C\&C board has been 
delivered and has been successfully tested. In parallel the firmware for the C\&C board is 
being developed.

\section{DAQ software}
The most important requirement on the DAQ software for the calorimeters is its scalability. 
In order to minimise risks and to shorten development time a well-established software framework 
will be used. DOOCS\,\cite{DOOCS} was found to be suitable for our DAQ needs. 

To adapt DOOCS to the requirements of our DAQ software the Equipment Name Server 
(ENS) has been used to integrate all DAQ components into one system. ENS provides 
services for naming resolution and RPC communication. In the naming structure of ENS
all the functionalities and properties for each device have
to be classified and defined. As seen in Fig.\,\ref{fig_daqoverview}, the ODR is the 
first hardware layer which can communicate with DOOCS. Therefore the software for the ODR has been 
the starting point of the DAQ software development. 

The DOOCS client provides a list of commands and configurable parameter with which the ODR devices can be managed. Configuration data and files are used when the system starts and runs. A prototype for the ODR DOOCS client has already been shown. To steer the whole system with the DOOCS control software several further developments are in progress.

\section{Conclusion and Outlook}
The DAQ system for the EUDET detector will be a hierarchical DAQ system which 
minimises the space needed on the detector and uses commercial components to minimise 
the cost of the DAQ system. In order to make the development for the test system 
independent of the detector type tested a DIF board will make the readout generic. 
An LDA card will concentrate the data from the DIF and send it off to the ODR which 
is located off the detector in a control room. The task of the ODR is to store the 
data.  All components have in addition the task to send configuration data and 
control messages to the detector and all other DAQ components upstream. At the time 
of writing each hardware component of the DAQ system exists and has been tested; the 
individual components now need to be integrated into a complete system. Current tests 
have shown that the DAQ software is able to integrate the ODR into the DAQ system 
based on the existing software framework DOOCS and manage successful communication
and control tests. The goal is to have the whole DAQ system ready for a test 
beam in 2009.

\section{Acknowledgements}
We would like to acknowledge the support of the Commission of the European 
Communities under
the 6th Framework Programme "Structuring the European Research Area",
contract number RII3-026126. 


\begin{footnotesize}

\end{footnotesize}



\begin{thebibliography}{99}
\bibitem{TESLA TDR} TESLA Technical Design Report, Part IV, A Detector for TESLA, T. Behnke et al., 2001
\bibitem{EUDET} EUDET - Detector R\&D towards the International Linear Collider, EUDET homepage:\\ 
{\tt http://www.eudet.org/}
\bibitem{HDMI} HDMI- High Definition Multimedia Interface Consortium homepage:\\ 
{\tt http://www.hdmi.org}
\bibitem{Enterpoint} Enterpoint, FPGA and ASIC Design, homepage: \\ 
{\tt http://www.enterpoint.co.uk}
\bibitem{PLDA} PLDA homepage: \\ {\tt http://www.plda.com/index.php}
\bibitem{DOOCS} DOOCS, Distributed Object Oriented Control System homepage:\\ 
{\tt http://tesla.desy.de/doocs/doocs.html}
 Linear Collider''.
\end{thebibliography}
\end{document}